\documentclass[pra,letterpaper,twocolumn,showpacs,superscriptaddress,floatfix]{revtex4}

\usepackage{graphicx}

\begin{document}

\title{All-optical cooling of Fermi gases via Pauli inhibition of spontaneous emission}

\author{Roberto Onofrio}

\affiliation{\mbox{Dipartimento di Fisica e Astronomia ``Galileo Galilei'', Universit\`a  di Padova, 
Via Marzolo 8, Padova 35131, Italy}}

\affiliation{\mbox{Department of Physics and Astronomy, Dartmouth College, 6127 Wilder Laboratory, 
Hanover, NH 03755, USA}}

\begin{abstract}
A technique is proposed to cool Fermi gases to the regime of quantum degeneracy based 
on the expected inhibition of spontaneous emission due to the Pauli principle. 
The reduction of the linewidth for spontaneous emission originates a corresponding 
reduction of the Doppler temperature, which under specific conditions may give 
rise to a runaway process through which fermions are progressively cooled.
The approach requires a combination of a magneto-optical trap as a cooling 
system and an optical dipole trap to enhance quantum degeneracy. 
This results in expected Fermi degeneracy factors $T/T_F$ comparable to the lowest 
values recently achieved, with potential for a direct implementation in optical lattices. 
The experimental demonstration of this technique should also indirectly provide a macroscopic 
manifestation of the Pauli exclusion principle at the atomic physics level.    
\end{abstract}


\maketitle

\section{Introduction}

One of the current goals of ultracold atom physics is to reach deeper quantum degeneracy for 
Fermi gases, which could enable the study of new phases of matter otherwise prevented by 
thermal effects \cite{Bloch,McKayDeMarco,Lewenstein}. The adaptation of evaporative cooling 
techniques so successful in the case of Bose gases to Fermi gases has been proven to be 
challenging, and so far the best degeneracy factors achieved have been in the 
$T/T_F \simeq 0.03-0.05$ range \cite{Shin,Partridge}. 
Physically, the origin of this plateau is rooted in the intrinsic difference between bosons and fermions. 
For the former, a cooling technique based on many-body effects such as evaporative cooling 
seems favourable as bosons tend to bunch in the same region of phase space and energy. Fermions instead 
manifest the opposite behavior, hence the lack of spatial overlap creates increasing difficulty as  
the degenerate regime is approached, further creating peculiar sources of heating \cite{Timmermans}.  
Limits to the efficiency of dual evaporative cooling between two distinguishable components of a 
Fermi gas have been discussed \cite{Crescimanno,Holland} and found to be in line with previous 
experimental findings \cite{DeMarco}. 
A compromise may consist in using sympathetic cooling of Fermi gas via Bose gases but, as 
noticed in \cite{Truscott}, the diminished heat capacity of a Bose gas at the onset 
of Bose-Einstein condensation makes sympathetic cooling increasingly inefficient. 
Although techniques to mitigate this limitation using species-selective traps have been proposed 
\cite{Onofrio,Presilla} and implemented in the Bose counterparts \cite{Catani,Baumer}, their 
broad application to reach quantum degeneracy in Fermi-Bose mixtures remains pending, with the 
notable exception of a very recent report on Yb-Rb mixtures \cite{Porto}.

Considering that a decade-long effort has not yet resulted in appreciable progress in Fermi cooling, one 
may ask whether it is possible to envision other cooling techniques which may transform the shortcomings 
of a Fermi gas into assets. In this paper, we discuss such a possibility by exploiting a well-known 
{\em individual} cooling technique, Doppler cooling in a Magneto-Optical Trap (MOT), and a 
{\em collective} effect unique of Fermi systems, the narrowing of the spontaneous emission linewidth. 
The analysis is first carried out on an atomic gas confined in a magneto-optical trap to illustrate 
the basics of the effect, as discussed in Sec. II. The limitations of Doppler cooling for fermionic 
isotopes of alkali-metal atoms brings to a more elaborated setup consisting in a hybrid 
configuration in which cooling is still ensured by the MOT, and trapping is achieved by a stiff 
optical dipole trap (ODT), as discussed in Sec. III. In Sec. IV we focus on the limitations to 
the cooling capability due to heating sources, and we discuss estimates of heating due to 
Rayleigh scattering in the ODT, showing that this should not compromise the achievement of 
quantum degeneracies comparable to the ones currently achieved with other cooling techniques.
Advantages and drawbacks of the proposal are then discussed in the conclusions, with particular 
regard to its implementation in optical lattices.

\section{Cooling by progressive suppression of the spontaneous-emission linewidth}

It is well-known that in the strongly degenerate regime a Fermi gas is expected to manifest 
inhibition of spontaneous emission \cite{Helmerson,Imamoglu,Javanainen,Busch,Goerlitz,Shuve}. 
As discussed in \cite{Busch}, three energy scales compete in determining the inhibition: 
the Fermi energy $E_F$, the recoil energy $E_R=2\pi \hbar^2/(2m \lambda_{\mathrm{at}}^2)$ 
(with $\lambda_{\mathrm{at}}$ the wavelenght of the Doppler cooling cycling transition), 
and the average thermal energy $k_B T$, with $k_B$ denoting the Boltzmann constant. 
Under the restrictions dictated by the Pauli principle, single-particle states in 
the energy range $E_F- k_B T \leq E \leq E_F+k_B T$ are partially occupied, so 
they can still accept an atom which undergoes spontaneous emission provided that its 
motional energy (given on average by the recoil energy plus the always present bonus of 
thermal energy, {\it i.e.} $E_R+k_B T$) falls in the same energy range. 
If, however, the Fermi energy is made too large with respect to $E_R$ and $k_B T$, the 
transition becomes progressively unlikely, as the motional energy is not enough to allow 
hopping into the energy levels still partially occupied. 
This effect has been studied quantitatively via explicit evaluation of the matrix 
element with numerical techniques \cite{Busch}, and with a local density approximation \cite{Shuve}. 
With reference to this last approximation, if $\Gamma_0$ and $\Gamma$ denote the intrinsic spontaneous 
emission linewidth and the spontaneous emission linewidth including the effect of the Pauli principle, 
respectively, then the spontaneous emission suppression factor $S_{SE}=\Gamma/\Gamma_0$ is \cite{Shuve}
\begin{equation}
S_{SE} \simeq z^{-1} \exp(\xi^2) f_3(z e^{-\xi^2}),
\end{equation}
where $z=\exp{(\beta \mu)}$ is the fugacity of the Fermi gas, with $\beta=(k_B T)^{-1}$ its inverse 
temperature, $\mu$ the chemical potential, $\xi=\sqrt{\beta E_R}$, and $f_3$  belongs to the set 
of functions defined in general in terms of the Fermi integral  
\begin{equation}
f_\nu(y)=\frac{1}{\Gamma(\nu)} \int_0^{\infty} dx \frac{x^{\nu-1}}{\exp(x/y)+1}, 
\end{equation} 
evaluated in the specific case of $\nu=3$, where $\Gamma(\nu)$ is the gamma function.  

\begin{figure}[t] 
\begin{center}
{\includegraphics[width=\columnwidth]{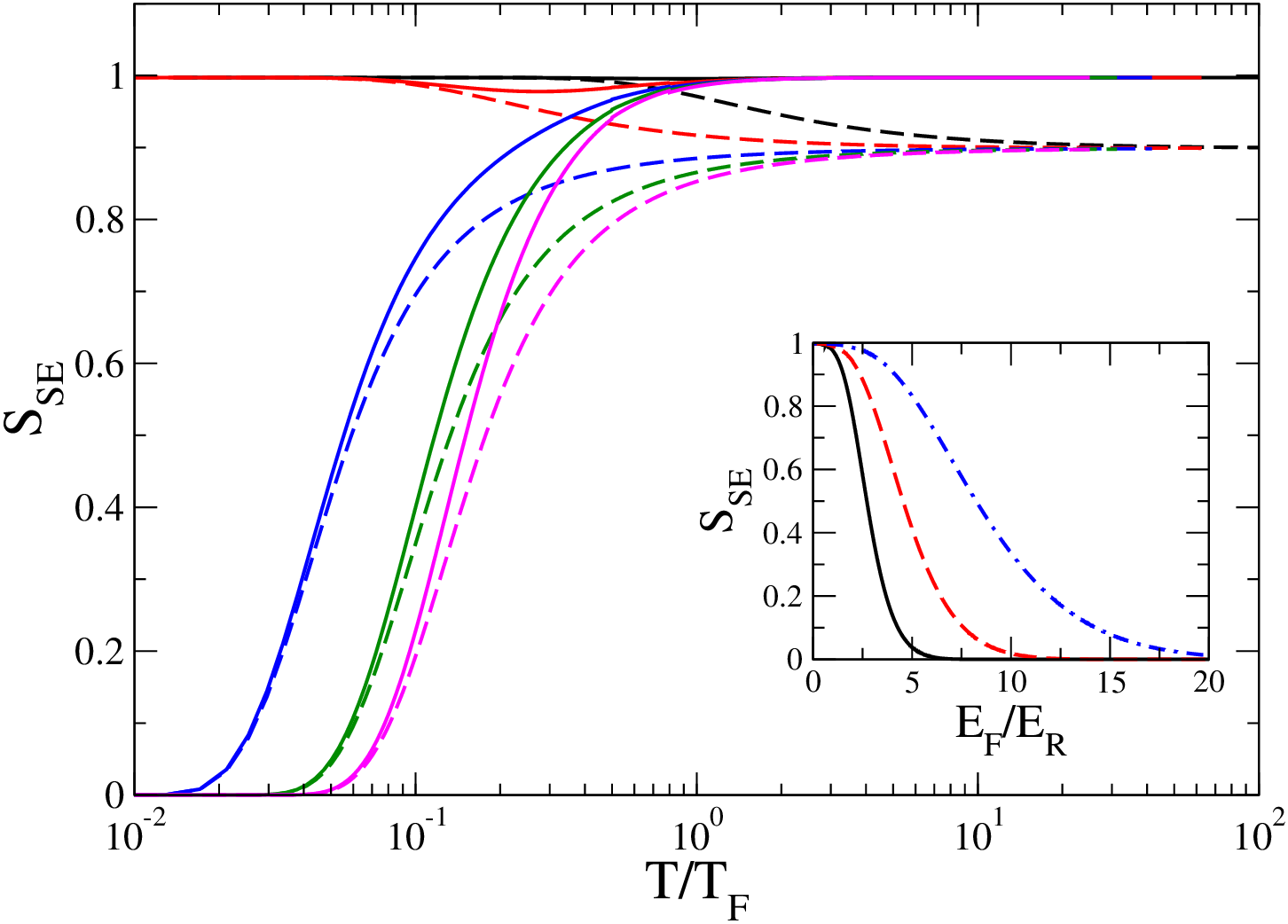}}
\caption{(Color online) Inhibition of the spontaneous emission for a Fermi gas. 
The spontaneous emission suppression factor $S_{SE}$ is plotted versus the degeneracy factor $T/T_F$ 
for different values of the Fermi energy in units of the recoil energy, from top to bottom 
curves, $E_F/E_R=0.4$ (black), $E_F/E_R=0.8$ (red), $E_F/E_R=1.2$ (blue ), $E_F/E_R=1.6$ (green), 
and $E_F/E_R=2.0$ (magenta). The continuous lines are evaluated assuming a temperature-dependent 
chemical potential, while the dashed lines are obtained by using the approximation $\mu= E_F$ 
valid only at low $T/T_F$ (see also Fig.~1 of \cite{Shuve}).   
In the inset the complementary dependence of $S_{SE}$ upon the $E_F/E_R$ ratio at constant 
temperature is shown for values $k_B T/E_R=0.5$ (black, continuous), $k_B T/E_R=1$ 
(red, dashed), and $k_B T/E_R=2$ (blue, dot-dashed).}
\label{Paulifig1}
\end{center}
\end{figure}

The dependence of $S_{SE}$ on the degeneracy factor $T/T_F$ is shown through the continuous 
lines in Fig.~1, for various values of the ratio $E_F/E_R$. The chemical potential of the 
Fermi gas has been evaluated at all temperatures with a function interpolating between the 
two extreme limits of the Sommerfeld expansion, 
\begin{equation}
\mu=E_F [1-(\pi k_B T/(\sqrt{3}E_F))^2], 
\end{equation}
and the high-temperature limit 
\begin{equation}
\mu=-k_B T \ln[6(k_B T/E_F)^3].
\end{equation}

The dashed lines in Fig.~1 are instead evaluated through the Sommerfeld approximation with 
the chemical potential equal to the Fermi energy, $\mu=E_F$, only valid at low $T/T_F$. 
The expected failure of this approximation at high temperature is manifested through an asymptotic 
value of $S_{SE} \simeq 0.9$ for $T \rightarrow \infty$ and more optimistic values of the 
suppression factor at any finite temperature.
The curves show that inhibition of spontaneous emission is ineffective at low values of 
$E_F/E_R$, and that it becomes progressively less sensitive to $E_F/E_R$ when this ratio 
becomes much larger than unity, an effect corroborated by showing its explicit dependence 
as in the inset of Fig.~1. 
\begin{figure*}[t] 
\begin{center}
{\includegraphics[width=0.48\textwidth]{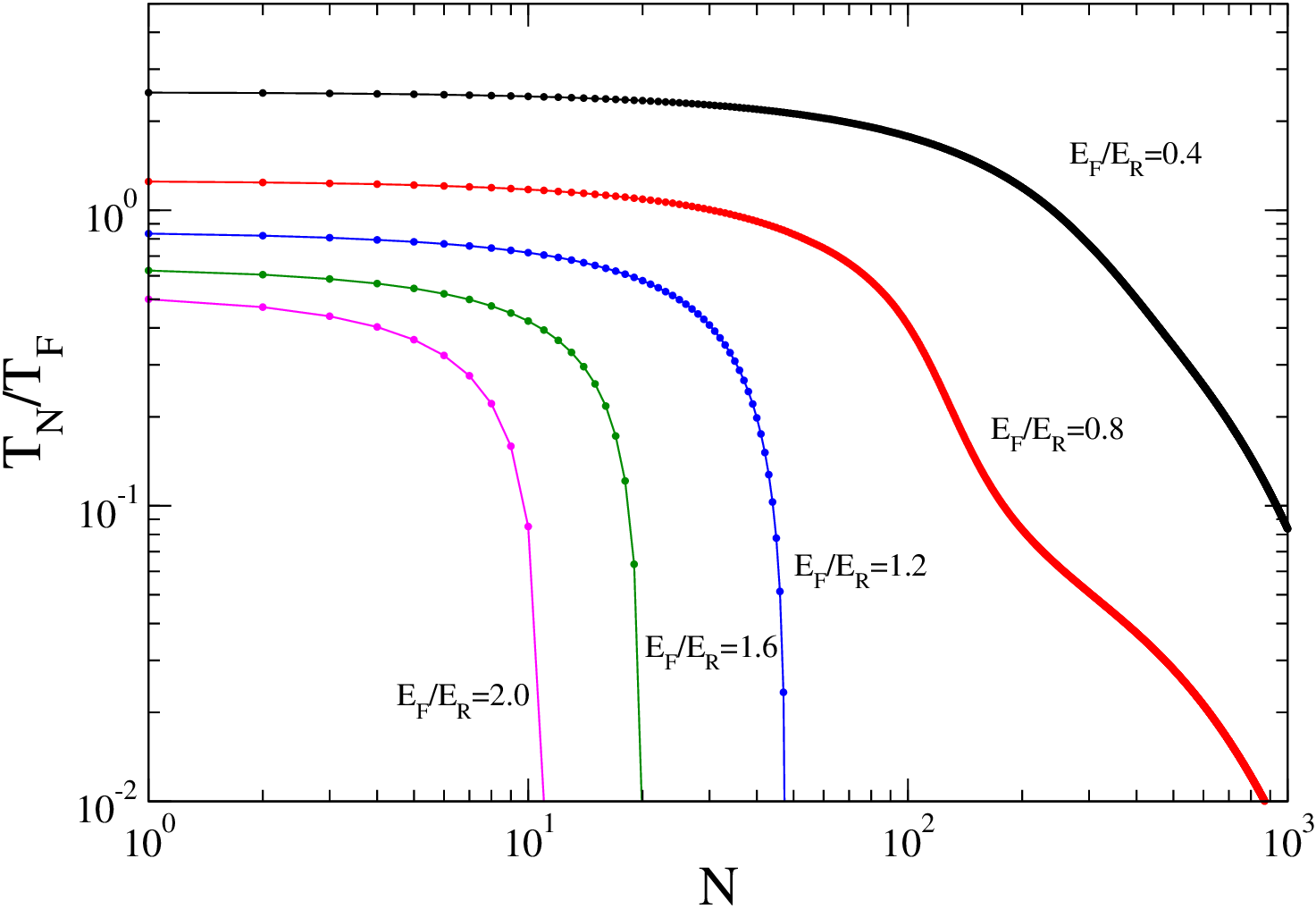}}
\hspace{0.2in}
{\includegraphics[width=0.48\textwidth]{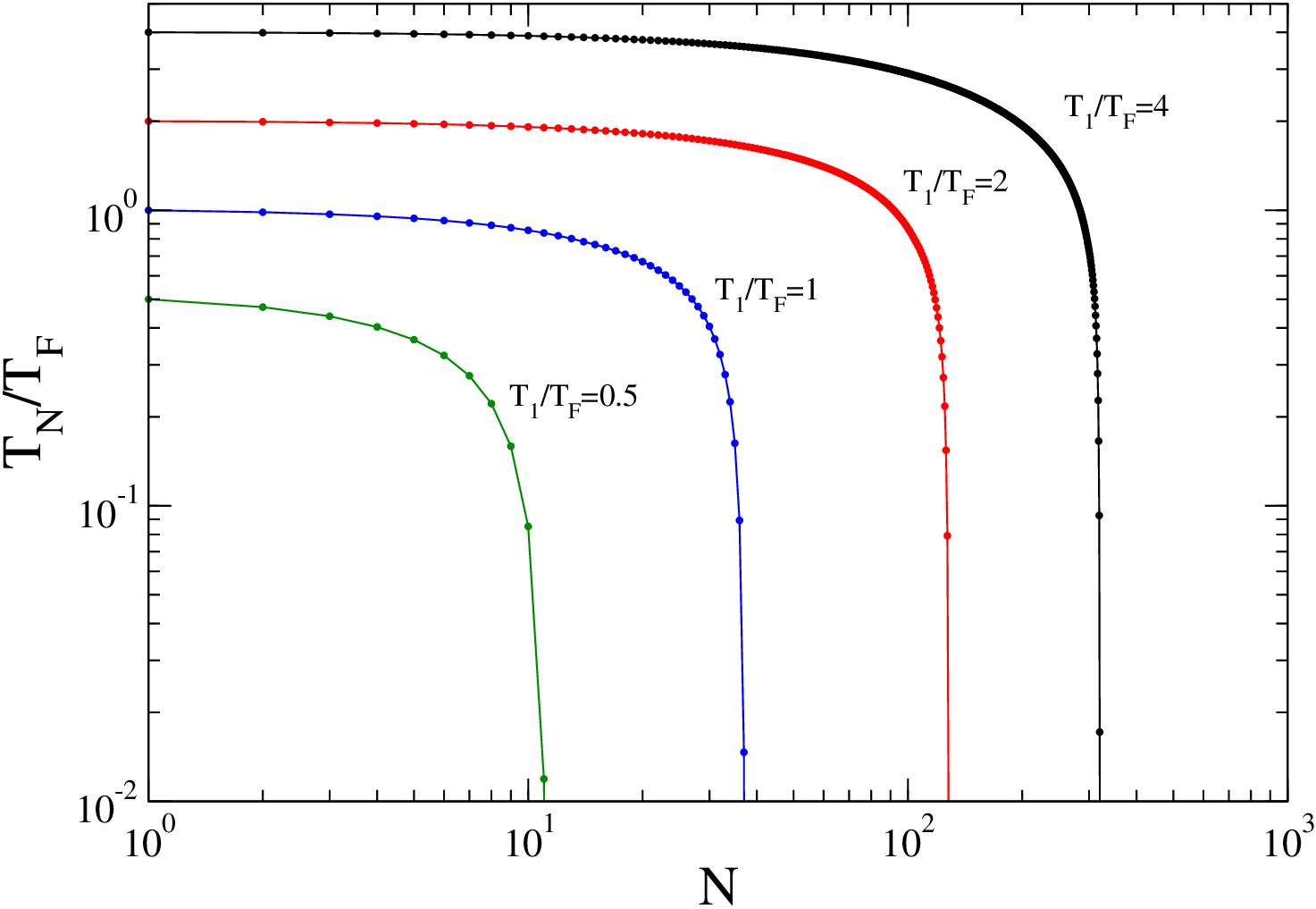}}
\caption{(Color online) Pauli cooling dynamics as based on the iterative equation Eq. (5). 
(a) Degeneracy factor $T_N/T_F$ versus number of iterations for 
the same values of the $E_F/E_R$ ratio used in Fig.~1. The initial temperature is 
assumed to be equal to the recoil temperature, as expected at the end of a MOT cooling 
protocol also exploiting a stage with optical molasses. 
(b) Degeneracy factor $T_N/T_F$ versus number of iterations for $E_F/E_R=2$ and different 
initial temperatures $T_1$, with the curve at the bottom (for $T_1/T_F=0.5$) corresponding to an 
initial temperature equal to the recoil temperature, $T_1=E_R/k_B$.} 
\label{Paulifig2}
\end{center}
\end{figure*}

The analysis of the plots suggests that, if the initial temperature is such that there 
is already some degree, even if small, of  quantum degeneracy, {\it i.e.} $T \lesssim T_F$, 
the linewidth $\Gamma$ should become slighty smaller than its intrinsic value $\Gamma_0$, and 
consequently the Doppler limit temperature $T_D = \hbar \Gamma/2k_B$ would be shifted to a smaller 
value, which also implies that the lowest temperature achieved during Doppler cooling will become 
smaller. This in turn implies a smaller degeneracy factor $T/T_F$, which implies a smaller 
$\Gamma$ and then again a smaller Doppler limit temperature. This loop will continue until the 
Doppler temperature becomes of the same order of magnitude of the recoil temperature.

This iterative dynamics can be made explicit with a simple recursive relationship in which each 
time step is of the order of the spontaneous emission lifetime $\tau=1/\Gamma=1/(S_{SE}\Gamma_0)$,
and we assume that equilibration timescales are comparable to this parameter.
The temperature of the atomic cloud after the N-th time step is then
\begin{equation}
T_{N+1}=S_{SE}(T_N/T_F)T_{N}, 
\end{equation}
where we indicate the explicit dependence on $T_F$ even if for now we consider a situation in 
which $E_F$ is kept constant. It is clear from this simple relationship and the plots in Fig.~1 
that there is a fixed point corresponding to zero temperature. In Fig.~2a we show the 
dynamics due to the iterative mapping of Eq. (5) for a Fermi gas precooled to the recoil limit 
temperature, that is, with initial temperature equal to $E_R/k_B$. 
Due to the highly nonlinear behavior of the iterative map, the temperature becomes extremely 
sensitive to the initial value of $E_F/E_R$ and the number of iterations. 
The fastest cooling trajectory shown in Fig.~2a occurs for $E_F/E_R=2$, in which a dozen 
of iterations appear to be enough to reach $T/T_F \leq 10^{-2}$.
Conversely, about $10^3$ iterations are required to reach $T/T_F\simeq 0.1$ if 
$E_F/E_R=0.4$, which is also aggravated by the fact that the actual time step is 
not uniform, but depends itself upon the degree of Fermi degeneracy. 
Indeed, being the spontaneous lifetime inversely proportional to the transition 
linewidth, the actual times for cooling and equilibration of a more degenerate 
cloud becomes proportionately longer, and the equally spaces steps in the horizontal 
axis of Fig.~2a should be replaced by time intervals of amplitude $\tau_N=1/(S_{SE}(T_N/T_F)\Gamma_0)$.
In Fig.~2b we show the complementary dependence of $T/T_F$ upon the number of iterations keeping 
constant the $E_F/E_R$ ratio and varying the initial temperature, giving informations on the 
sensitivity of the iterative scheme upon initial states with temperature higher than the recoil 
temperature. Alternatively, the iterative relationship in Eq. (5) can be easily converted into 
a differential equation keeping in mind that the timescale for cooling is of the order of the 
spontaneous emission lifetime,
  
\begin{equation}
\frac{dT}{dt} \simeq \frac{T_{N+1}-T_N}{\tau_N}=-S_{SE}(1-S_{SE})\Gamma_0 T, 
\end{equation}
in which an effective cooling timescale may be identified as 
$\tau_{\mathrm cool}=[S_{SE}(1-S_{SE})\Gamma_0]^{-1}$ depending on the degeneracy 
parameter $T/T_F$ and then on time, confirming that the dynamics is highly nonlinear. 
The maximal cooling speed is achieved in the intermediate regime $S_{SE} \simeq 1/2$, while 
it tends to zero both in the case of $S_{SE}$ close to unity -- {\it i.e.} far from quantum degeneracy, 
when the linewidth suppression is not present -- and when $S_{SE}$ is close to zero, {\it i.e.} when 
the spontaneous lifetime tends to infinity slowing down the cooling process. In Fig.~3 we show 
the implications in terms of the time dependence of the degeneracy parameter for the same 
iterative scheme discussed in Fig.~2a, and the spontaneous emission lifetime of the dominant 
transition $2s_{1/2}-2p_{3/2}$ in ${}^6$Li, ($\lambda_{\mathrm{at}}=671$ nm, $\Gamma=2\pi \times 5.9$ MHz). 
While all curves show $T/T_F \rightarrow 0$ as $t \rightarrow \infty$, their behavior at 
finite time is different, with the appearance of an optimal, intermediate value of 
$E_F/E_R$ leading to maximum cooling, a situation of interest in any realistic scheme, in 
which obviously a nonzero cooling power is required.

\section{Cooling in a MOT assisted by an optical dipole trap}

The previous analysis has shown that the initial degeneracy parameter $T/T_F$ is crucial 
to ignite the expected runaway process, as specifically seen in Fig.~2b. 
In applications to specific Fermi gases, with ${}^6$Li and ${}^{40}$K being the only 
alkali-metal stable or long lifetime fermionic isotopes, the temperature is initially 
limited by the Doppler temperature itself due to the unresolved hyperfine structure 
preventing efficient optical molasses cooling. This issue is generally not present 
in alkaline-earth metal fermionic species, for which the starting point can be a 
sub-Doppler temperature, which in the presence of a degeneracy parameter $T/T_F$ small enough 
can start the runaway process until the Doppler temperature reaches the recoil temperature. 
We deliberately address in the following the less favourable situation of ${}^6$Li and 
${}^{40}$K, also due to their widespread availability in several laboratories. 
The cooling process can be made more efficient in spite of the absence of genuine sub-Doppler 
cooling stages by both increasing its speed, related to the $E_F/E_R$ ratio as visible in Fig.~2a, 
and decreasing the $T/T_F$ ratio. This is simultaneously achieved by adding an ODT sharing the 
same potential energy minimum as the MOT, which will increase the Fermi energy thereby 
decreasing $T/T_F$ while increasing $E_F/E_R$. 
In this condition, the achievement of a temperature close to the recoil temperature is no 
longer a strong limitation, as far as we are concerned with getting the minimum $T/T_F$ and 
not minimum temperatures in an absolute sense. Fortunately, most of the interesting phase 
transition physics depends on the entropy, which in turn depends on the $T/T_F$ ratio. 

\begin{figure}[t] 
\begin{center}
{\includegraphics[width=0.48\textwidth]{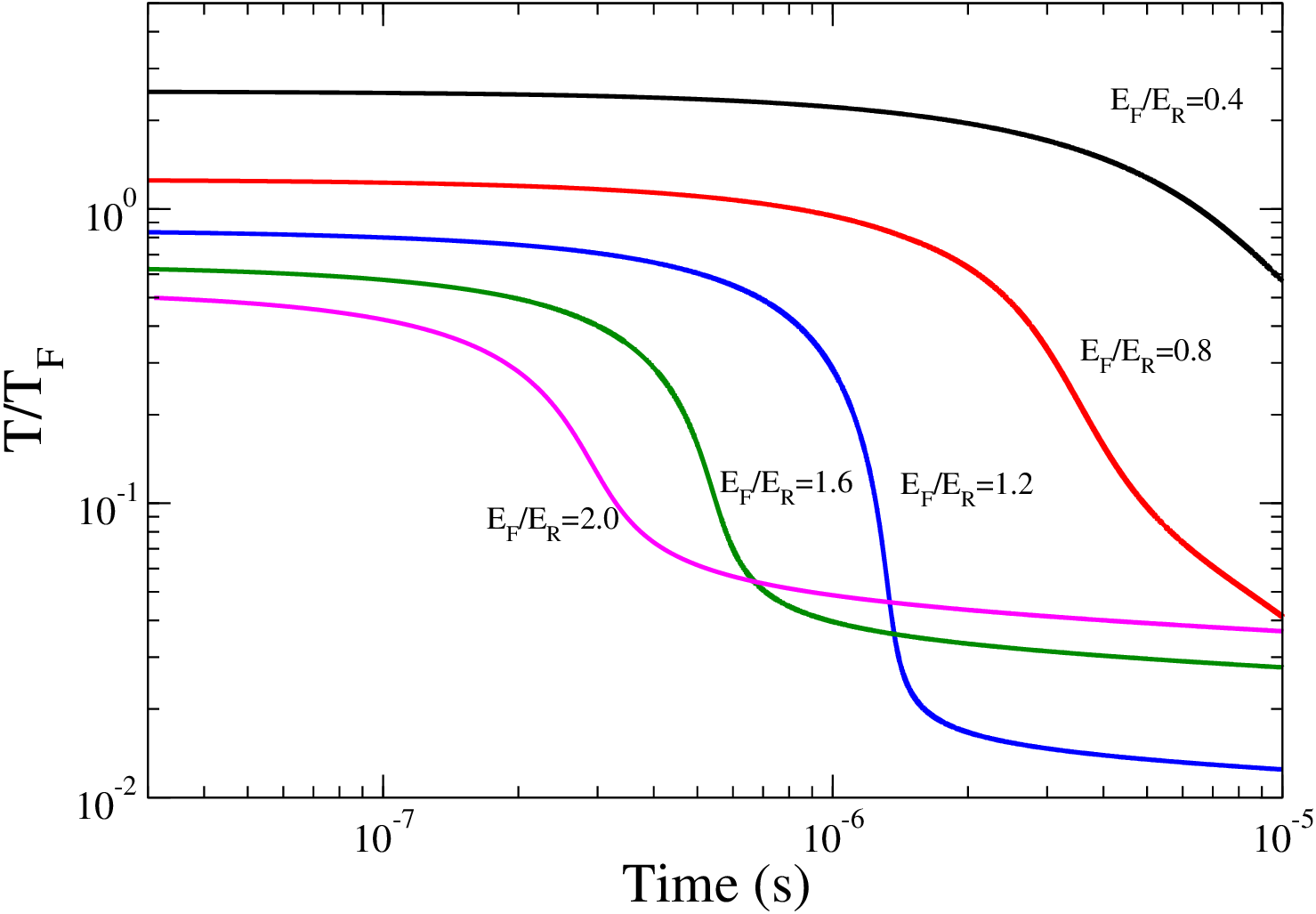}}
\caption{(Color online) Same case as in Fig.~2a but with the actual time in abscissa, 
with the time evolution ruled by Eq. (6). Although it seems that even for moderate 
values of the $E_F/E_R$ ratio the degeneracy factor decreases, this occurs at the price 
of a much slower timescale with respect to the case of large $E_F/E_R$ ratios. 
Notice that too large $E_F/E_R$ ratios imply large spontaneous lifetimes achieved at 
early times, which results in precocious slow down of the cooling dynamics, as visible  
especially for the $E_F/E_R=2$ case.} 
\label{Paulifig3}
\end{center}
\end{figure}

The coexistence of a MOT only utilized for Doppler cooling and an ODT for stiff trapping is not 
an issue provided that the laser wavelength for the ODT is judiciously chosen to avoid any ac 
Stark shift of the two levels realizing the cycling cooling transition in the MOT, as achievable 
by using the so-called magic wavelengths \cite{Herold,Safronova}. Production of large degenerate 
samples of ${}^6$Li has been recently reported in \cite{Burchianti}, to which we also refer for  
discussions of experimental details, in which the authors have reported efficient Doppler and 
molasses cooling in the presence of the strong light field of the ODT. 

\begin{figure}[t]
\begin{center}
{\includegraphics[width=\columnwidth]{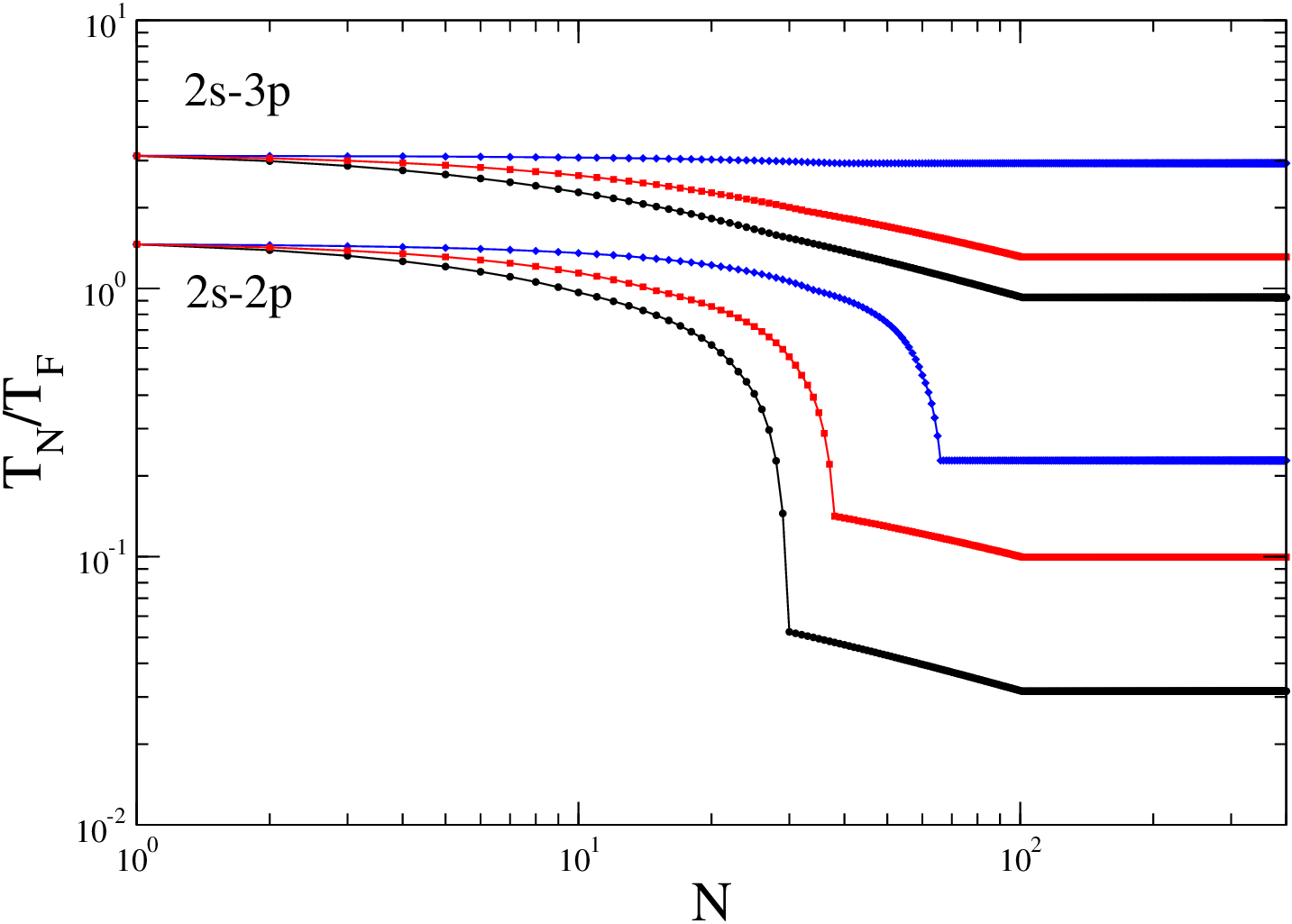}}
\caption{(Color online) Pauli cooling of a Fermi gas in a hybrid combination of a 
magneto-optical trap and an optical dipole trap. The three bottom curves show the 
effect of the optical dipole trap for trapping and cooling $10^5$ ${}^6$Li atoms 
in the three cases of constant laser power (blue diamonds, dot-dashed) of $P=$10 W, 
and a linear ramp starting from the initial time until $10^2$ iterations with the 
same initial value of 10 W and final value of 50 W (red squares, dashed) and 100 W 
(black circles, continuous), with a beam waist $w=20 \mu$m. 
The initial temperature of the atoms is assumed to be the Doppler temperature for the 
$2s_{1/2}-2p_{3/2}$ cyclic transition, $T_D=140~ \mu$K, and the laser wavelength is assumed to 
be $\lambda_{\mathrm{las}}=$930.3 nm, minimizing the Stark shifts induced by the ODT. 
The three top curves are relative to the case of a $2s_{1/2}-3p_{3/2}$ cyclic transition with the 
initial temperature assumed to be the corresponding Doppler temperature of $T_D=18~ \mu$K,   
a laser wavelength of $\lambda_{\mathrm{las}}=$1070.73 nm, and the same laser power trajectories as 
for the $2s_{1/2}-2p_{3/2}$ cyclic transition. 
Due to the weaker optical confinement as a consequence of the smaller linewidth $\Gamma$, the 
latter cyclic transition is not favourable despite a smaller initial Doppler temperature.}  
\label{Paulifig4}
\end{center}
\end{figure}

We focus here on the case of ${}^6$Li, and a crossed ODT produced by 
a laser working at one magic wavelength for the relevant ${}^6$Li cyclic transition. 
The trapping frequencies along the three axes may be written as 
\begin{equation}
\omega_x=\omega_y=\frac{\omega_z}{\sqrt{2}}=\left(\frac{\hbar \alpha P}{\pi m w^4}\right)^{1/2}, 
\end{equation}
where $P$ is the instantaneous laser power, $w$ the waist of the laser beam, 
$\alpha=\Gamma^2/\{[1/(\Omega_{\mathrm{at}}-\Omega_{\mathrm{las}})]I_{\mathrm{sat}}\}$, where 
$I_{\mathrm{sat}}=\hbar \Omega_{\mathrm{at}}^3 \Gamma/(12 \pi c^2)$ is the saturation intensity, 
with $\Omega_{\mathrm{at}}=2\pi c/\lambda_{\mathrm{at}}$, 
$\Omega_{\mathrm{las}}=2\pi c/\lambda_{\mathrm{las}}$, 
$\lambda_{\mathrm{at}}$ and $\lambda_{\mathrm{las}}$ the atomic transition and laser wavelength, 
respectively. The case of the crossed ODT, although more difficult to achieve 
due to the need of focusing two laser beams on the MOT center, is the 
closest to isotropy and avoids excessive spatial mismatching with respect to 
the pre-existing shape of the atomic cloud in the MOT. Also, in order to avoid sudden 
heating of the atomic cloud, the initial laser power should be chosen appropriately to 
allow for a smooth transition from the MOT trapping to the ODT trapping, thereby the need 
to start with a moderate value of the laser power then progressively ramped up.
The linear dependence of the Fermi energy on the average trapping frequency 
$\omega=(\omega_x \omega_y \omega_z)^{1/3}$, as $E_F = 1.82 \hbar \omega N_F^{1/3}$, and 
the fact that the average trapping frequency scales as $P^{1/2}$ implies a scaling of 
the degeneracy parameter as $T/T_F \propto P^{-1/2}$.

The effect of the ODT on the iterative scheme is shown in Fig.~4 for various laser 
powers and two different cyclic transitions. The case of the $2s_{1/2}-3p_{3/2}$ cyclic 
transition seems apparently more suitable, since in this case the spontaneous emission 
linewidth is narrower, and in fact slightly larger than the recoil temperature, making the 
initial conditions more appealing for Pauli cooling. However, due to the dependence of the 
trapping frequencies on the linewidth and the saturation intensity, the initial confinement 
is weaker than in the alternative situation of the $2s_{1/2}-2p_{3/2}$ cyclic transition, if the comparison 
is made for the same laser power and waist. This weaker trapping slows down so much the cooling 
process that, as seen in Fig.~4, the gain in $T/T_F$ is only due to the ramping of the laser power 
and the increase in $T_F$, rather than the intrinsic decrease of $T$. Conversely, in the case of 
the $2s_{1/2}-2p_{3/2}$ transition, the cooling is more effective and reaches the recoil limit well 
before the laser power is settled at a constant value after $10^2$ iterations, at which point 
further gains in the $T/T_F$ ratio are only achieved by increasing $T_F$.  The case of a constant 
and shallow ($P$=10 W) optical trap is also shown, to emphasize both the gain in speed and in the 
achievable mimimum $T/T_F$ ratio by using time-dependent laser powers.   

\section{Limitations to cooling due to heating sources}

An obvious limitation in the use of a high power ODT is due to the concurring heating 
rate by off-resonant Rayleigh scattering. Also, higher densities of the Fermi gas will make 
molecule formation more likely, and with it a faster decay of the atomic gas through three-body 
collisions and related processes. In Fig. 5 we show the impact on the cooling scheme of two 
additional limiting factors. First, the ODT introduces a heating source due to Rayleigh scattering.  
Although the off-resonance nature of the laser beam should limit this heating source, the 
effect may affect significantly the cooling efficiency in the latest stages due to the linear 
dependence of the specific heat of a Fermi gas on the $T/T_F$ ratio, which makes the atomic gas 
more susceptible to any heating source. Second, as anticipated in Sec. II, in a realistic modeling the 
iteration steps should be replaced by the actual time interval occurring for each step, which 
depends on the spontaneous lifetime and the suppression factor $S_{SE}$. This is a further limitation 
to the cooling scheme since it slows down the dynamics, while Rayleigh scattering heating (although 
suppressed by quantum degeneracy) and any other source of heating are present. 
The specific heat per unit of atom has been evaluated interpolating between the degenerate 
regime and the Dulong-Petit limit with an exponential function $C_F=3k_B[1-\exp(-\pi^2k_BT/(3 E_F))]$.
The dramatic time dependence of the case with final power of 100 W shows that, once  
a deep Fermi degeneracy is reached, the time scale slows down significantly. 
This implies a large exposure to Rayleigh heating, further enhanced by the decreased specific 
heat -- thereby the sudden jump, in the next iteration, to much higher degeneracy factors.

\begin{figure}[t]
\begin{center}
{\includegraphics[width=\columnwidth]{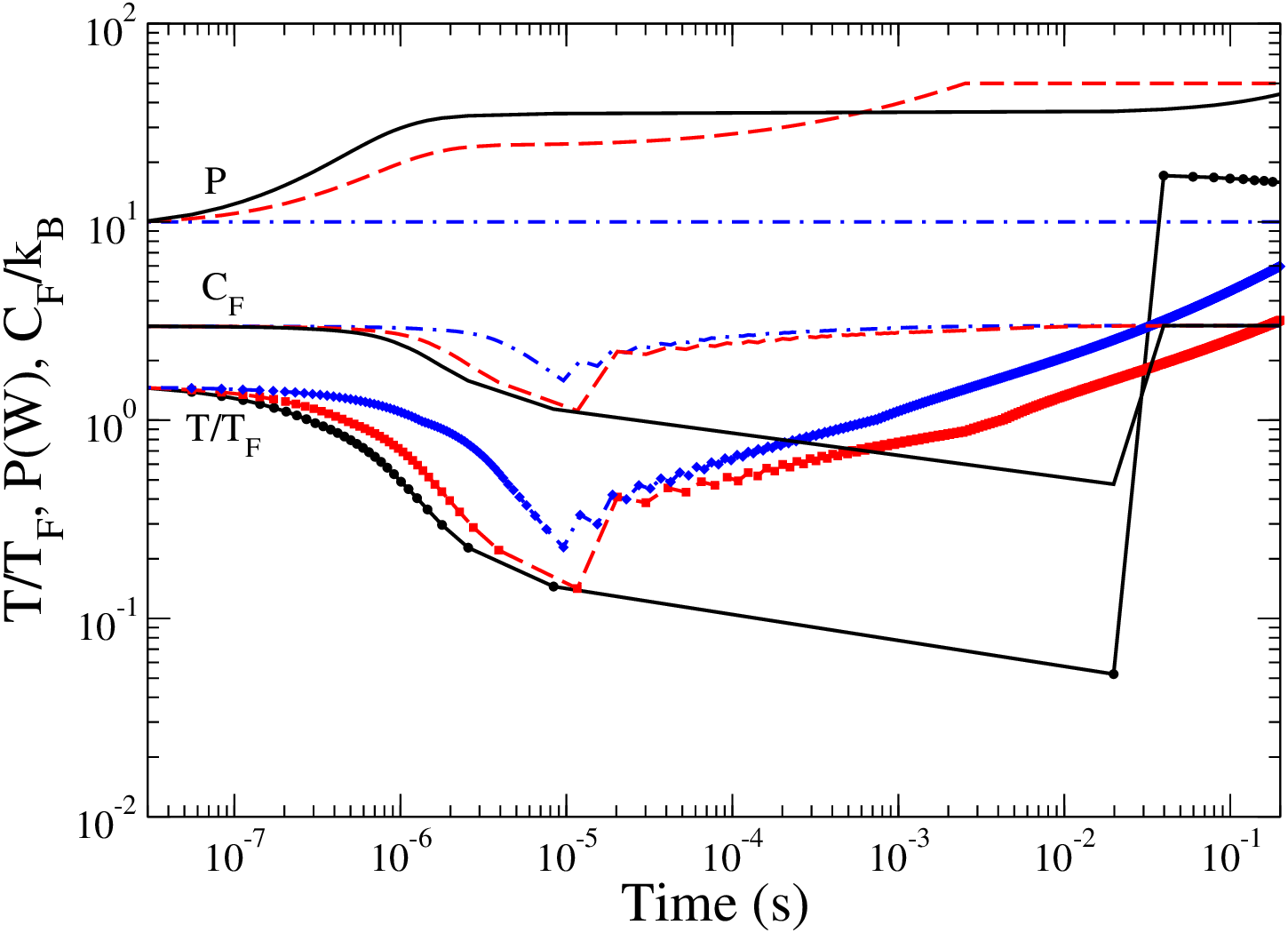}}
\caption{(Color online) Time dependence of the degeneracy factor $T/T_F$ in the presence of heating 
due to Rayleigh scattering from the optical dipole trap. The parameters and the protocols 
are the same as for the $2s_{1/2}-2p_{3/2}$ cyclic transition discussed in Fig. 3. 
Plotted are the time dependence of the laser power (top curves) for a constant power 
of 10 W (blue, dot-dashed), a linear ramp from 10 W to 50 W then kept constant (red, dashed), 
and a linear ramp from 10 W to 100 W then kept constant (black, continuous), the related 
specific heats in Boltzmann constant units (middle curves), and the resulting $T/T_F$ 
trajectories (bottom curves). }
\label{Paulifig5}
\end{center}
\end{figure}

The plots show that there is a minimum in the degeneracy factor $T/T_F$ achieved when the 
cooling and heating rates compensate each other. To avoid the detrimental stage occuring 
after this minimum, the laser power should be accordingly decreased, thereby allowing for 
an additional stage of evaporative cooling. 

The presence of a minimum achievable value for $T/T_F$ can also be evidenced by analytical 
estimates completing the differential equation Eq. (6) to include the heating source, 
\begin{equation}
\frac{dT}{dt} = \dot{T}_{\mathrm{cool}}+\dot{T}_{\mathrm{heat}}
=-S_{SE}(1-S_{SE})\Gamma_0 T + \frac{P_{\mathrm{heat}}}{C_F}, 
\end{equation}
where $P_{\mathrm{heat}}$ is the heating power, and it should be remarked that both 
$S_{SE}$ and $C_F$ depend upon the degeneracy factor $T/T_F$, yielding a highly nonlinear 
dynamics. The heating term is independent on temperature in the Dulong-Petit nondegenerate 
regime, $\dot{T}_{\mathrm{heat}} \simeq P_{\mathrm{heat}}/3k_B$, while increases in inverse 
proportion to the degeneracy factor in the fully degenerate regime, 
$\dot{T}_{\mathrm{heat}} \simeq P_{\mathrm{heat}}/(\pi^2 k_B) \times E_F/(k_BT)$. 
Therefore the most delicate stage for heating occurs at later times,  when the linearly 
decreasing specific heat makes the atomic temperature increasingly susceptible to the 
presence of heating sources. Likewise, as commented earlier the optimal cooling situation 
is achieved if the spontaneous scattering suppression factor is $S_{SE}=1/2$, {\it i.e.} when 
the system is in an intermediate quantum degeneracy regime, and in this situation 
$\dot{T}_{\mathrm{cool}}= -\Gamma_0 T/4$. Under this intermediate regime, the balance between 
cooling and heating rates becomes rather simple, leading to a rough estimate of the minimum 
degeneracy factor achievable if the limitation is due to heating sources as  

\begin{equation}
\left(\frac{T}{T_F}\right)_{\mathrm{min}} \simeq 2 \pi \left(\frac{P_{\mathrm{heat}}}
{\Gamma_0 E_F}\right)^{1/2}. 
\end{equation}

We consider the case of heating power due to residual Rayleigh scattering in the 
ODT alone, so $P_{\mathrm{heat}}=2 \gamma_{sc} \tilde{E}_R$, where 
\begin{equation}
\gamma_{sc}=\frac{\Gamma^3}{8}\left(\frac{1}{\Omega_{\mathrm{at}}-\Omega_{\mathrm{las}}}+
\frac{1}{\Omega_{\mathrm{at}}+\Omega_{\mathrm{las}}}\right)^2
\left(\frac{\Omega_{\mathrm{las}}}{\Omega_{\mathrm{at}}}\right)^3 \frac{I}{I_{\mathrm{sat}}}.
\end{equation} 
is the scattering rate and $\tilde{E}_R=2\pi \hbar^2/(2m \lambda_{\mathrm{las}}^2)$ is the recoil 
energy due to the scattering of photons from the ODT laser beams. Notice that, due to the 
linear dependence on $\Gamma$ of $I_{\mathrm{sat}}$, the heating power depends quadratically on 
$\Gamma$ and then on the spontaneous suppression factor $S_{SE}$, making its contribution smaller 
in the degenerate regime, partially compensating for the larger temperature changes due to 
the linear drop of the specific heat of the Fermi gas. 

\begin{figure}[t]
\begin{center}
{\includegraphics[width=\columnwidth]{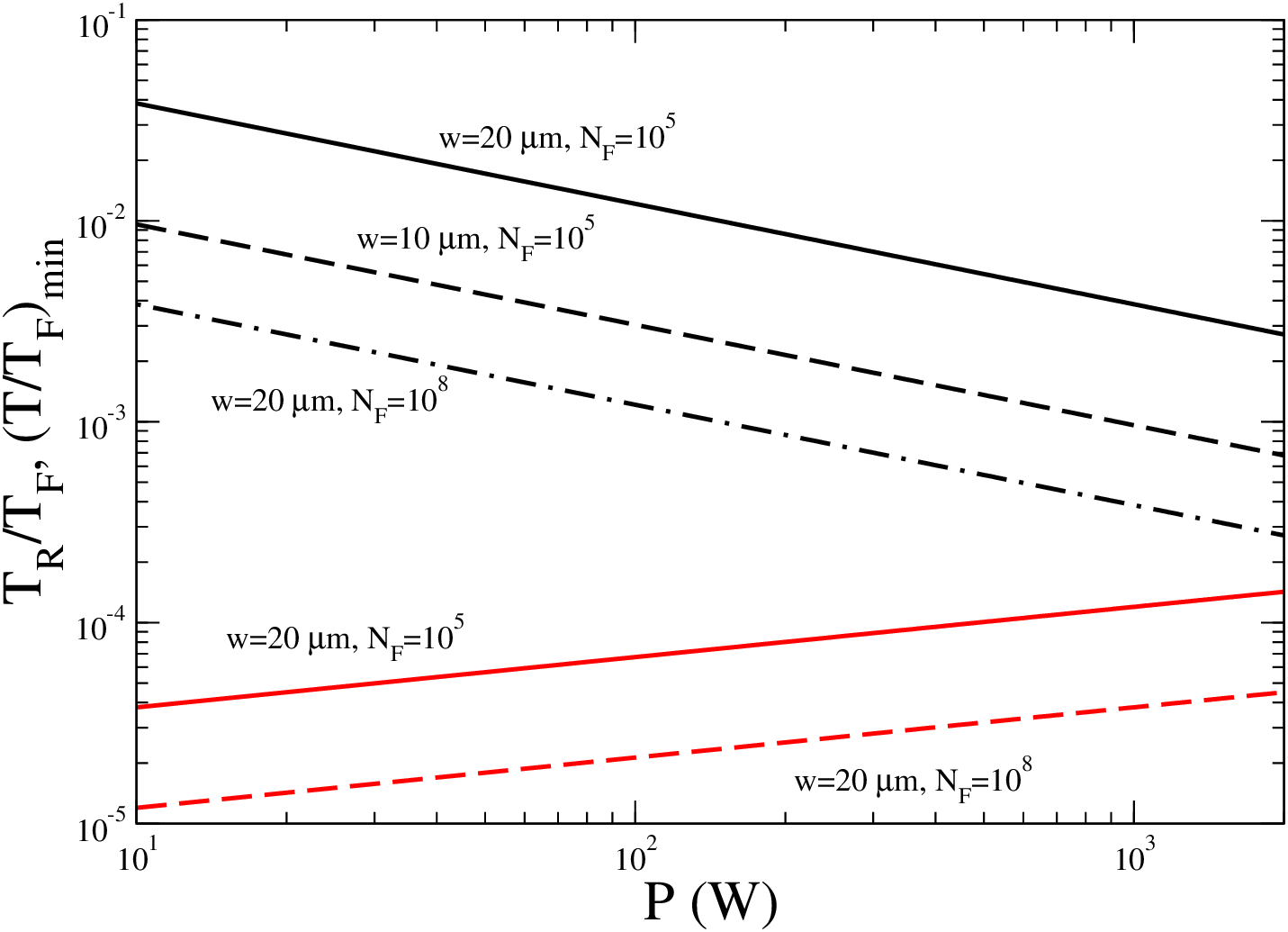}}
\caption{(Color online) Limits to the degeneracy factor $T/T_F$ arising from the recoil limit 
(three top black curves) and the heating and cooling rates estimated in the text through Eq. (9) 
(two bottom red curves), versus the laser power and for various values of the beam waist and 
number of fermions. The parameters are the same as for the $2s_{1/2}-2p_{3/2}$ cyclic transition 
discussed in Fig. 4.}
\label{Paulifig6}
\end{center}
\end{figure}
 
For the case of ${}^6$Li $2s_{1/2}-2p_{3/2}$ transition and power and waist of the ODT laser beams  
respectively of 10 W and 20 $\mu$m as in the example in Fig.~4, this gives rise to a lower 
bound of $k_BT/E_F \simeq 4 \times 10^{-5}$. This is not the actual limiting factor since the 
recoil limit $k_B T_R/E_F \simeq 4 \times 10^{-2}$, in line with what observed in Fig.~5, is 
larger by three orders of magnitude. Moreover, the bound expressed by Eq. (9) yields values of 
$T/T_F$ for which the initial hypothesis of $S_{SE}=1/2$ does not hold, as a visual inspection 
of Fig.~1 shows that for the estimated $T/T_F$ we get $S_{SE} << 1/2$.   
Therefore, the analytical estimate in Eq. (9) should be considered as a lower bound on the 
minimum achievable $T/T_F$ due to heating sources, and its value is also in comparing how 
the two limitations in $T/T_F$ depend upon the laser power, the waist of the beams, the number 
of fermions, as shown in Fig.~6. 
The dominance of the recoil limit is evident for any realistic value of the laser power at 
a moderately large waist of 20 $\mu$m and for $10^5$ fermions. The heating limit does not depend 
on the waist of the laser beam since both the scattering rate and the Fermi energy scale as an inverse 
square power of the waist, however larger number of fermions will increase $E_F$ with different 
scalings for the two bounds $E_R/E_F$ and $(T/T_F)_{\mathrm{min}}$ (proportional to $N_F^{-1/3}$ and 
to $N_F^{-1/6}$ respectively), making the limitation due to heating progressively more relevant. 
 
The numerical examples reported here are not fully optimized as they are chosen with an eye 
towards trapping performances already achieved in various laboratories. Given the high degree 
of nonlinearity in the protocol and the dependence on the relevant parameters, improvements by 
at least one order of magnitude with respect to the estimated degeneracy parameter, could be 
realistically achievable in specific configurations. The cooling timescale -- as visible in 
Fig.~5 -- seems short, at least for the specific case of the ${}^6$Li transition used as an example. 
However, more elaborated cooling cycles based upon ramping up and down repeatedly the laser power 
may extend the lifetime of the ultracold sample well beyond the millisecond scale. Considering the 
large trapping frequencies achieved in high-power ODTs, and the fact that the inverse of the 
chemical potential usually determines the timescale for building up strongly correlated phases, 
we do not expect the cooling timescale being an issue in the study of several phase transitions 
predicted so far in the ultracold regime. Even without fully optimizing the many parameters 
available, a final degeneracy factor $T/T_F$ below the $10^{-2}$ range seems within reach, as 
noticeable in particular looking at the limits in Fig.~6. 

\section{Conclusions}

In conclusion, we have discussed an experimental setup in which a peculiar feature 
of degenerate Fermi gases, the expected narrowing of the spontaneous emission linewidth, 
may be exploited as a cooling mechanism. 
The same advantages of all-optical cooling of atomic gases discussed elsewhere 
do apply here, without the loss of atoms as in dual evaporative cooling, the 
complication of simultaneous trapping of a Bose gas with large heat capacity as 
in sympathetic cooling, and with the possibility to fully exploit Feshbach resonances. 
The concurrent suppression of light scattering expected for a degenerate Fermi gas has also two 
added benefits. First, the reduced radiation pressure experienced by the atoms allows for higher  
atomic densities, and atomic heating by light scattering is suppressed in the strong confinement 
provided by the ODT, as experimentally demonstrated in an optical lattice \cite{Wolf}. 
Second, continuous and precision monitoring of the cloud temperature through 
off-resonant beams over the duration of the experiment \cite{Shuve} helps to mitigate another 
critical issue for degenerate Fermi gases, namely the lack of sensitive precision thermometry in 
the degenerate regime, a crucial point for the study of classical phase transitions \cite{McKayDeMarco}. 

The use of tight trapping conditions makes this cooling scheme less palatable for the 
application to optical lattices in which a degree of spatial homogeneity is often a prerequisite 
for creating and observing unconventional phases. However, the combination of this cooling mechanism 
with adiabatic expansions, such as the ones designed through frictionless cooling \cite{Chen,Torrontegui}, 
should mitigate this drawback. Moreover, a direct implementation of Pauli blocking schemes 
to further cooling in optical lattices seems feasible \cite{Sandner}. 

A full expansion of the proposal should also include a treatment of the dynamics of the atoms in a 
combined MOT-ODT configuration, as this gives rise to confinements rather different in various spatial 
regions, with a small, inner region experiencing stiff confinement and a large, outer region loosely 
confined, and not contributing to the cooling effect exploited in this scheme. 
The proposal could take advantage of efforts to produce a large volume ODT \cite{Mosk,Stellmer}, also 
complementing existing demonstrations of efficient all-optical cooling of atomic gases, either 
alkaline-earth metals \cite{Katori,Binnewies,Curtis,Loftus}, or fermionic isotopes of lithium 
and potassium \cite{Duarte,McKay,Sebastian} via intrinsically narrow transitions. 
The experiment described in \cite{Burchianti} seems the closest, both in the set up and 
parameters space, to a successful implementation of the technique proposed in this paper.
From a broader, fundamental physics standpoint, implementing this cooling technique should 
also yield a beautiful macroscopic demonstration of the Pauli principle, complementing 
recently reported evidences of microscopic character \cite{Omran}.

\vspace{0.5cm}

\begin{acknowledgments}
I am grateful to Lorenza Viola and Kevin C. Wright for a critical reading of the manuscript.
\end{acknowledgments}

\end{document}